\documentclass[aps,twocolumn,preprintnumbers,amsmath,amssymb,floatfix,footinbib,longbibliography]{revtex4-2}
\usepackage{graphicx}
\usepackage{longtable}
\usepackage{dcolumn}
\usepackage{bm}
\usepackage{float}
\usepackage{xcolor}
\usepackage{etoolbox}
\usepackage{comment}
\makeatletter
\patchcmd{\@citex}{\@citeb}{\csname b@\@citeb\endcsname\@extra@b@citeb}{}{}
\makeatother
\usepackage[normalem]{ulem}
\usepackage[colorlinks=true, urlcolor=blue, linkcolor=blue, citecolor=blue]{hyperref}
\usepackage[all]{hypcap}
\usepackage{hyperref}     
\hypersetup{
    colorlinks,
    citecolor=blue,
    linkcolor=blue,
    urlcolor=blue}
\newcommand{\1}{\c{c}}

\newcommand{\beq}{\begin{equation}}
\newcommand{\eeq}{\end{equation}}
\newcommand{\bea}{\begin{eqnarray}}
\newcommand{\eea}{\end{eqnarray}}

\begin{document}

\title{Chiral split magnons in metallic g-wave altermagnets: Insights from many-body perturbation theory}

\author{W. Beida$^{1,2}$}\email{w.beida@fz-juelich.de}
\author{E. \c{S}a\c{s}{\i}o\u{g}lu$^3$}\email{ersoy.sasioglu@physik.uni-halle.de}
\author{C. Friedrich$^1$}
\author{G. Bihlmayer$^1$}
\author{Y. Mokrousov$^{1,5}$}
\author{S. Bl\"{u}gel$^{1}$}
\affiliation{$^{1}$Peter Gr\"unberg Institut, Forschungszentrum J\"ulich and JARA, 52425 J\"ulich, Germany \\
$^{2}$ Physics Department, RWTH-Aachen University, 52062 Aachen, Germany \\
$^{3}$Institute of Physics, Martin Luther University Halle-Wittenberg, 06120 Halle (Saale), Germany \\
$^{5}$Institute of Physics, Johannes Gutenberg University Mainz, Mainz, Germany}
\date{\today}

\begin{abstract}

Altermagnets are a novel class of magnetic materials that bridge the gap between ferromagnets (FMs) and antiferromagnets (AFMs). A key feature is the non-degeneracy of magnon modes where spin splitting occurs, leading to chirality and direction-dependent magnon dispersions governed by symmetry. We explore this in metallic g-wave altermagnets (\(TPn\), where \(T\)= V, Cr; \(Pn\)= As, Sb, Bi) using density functional and many-body perturbation theories. We analyze the influence of pnictogen substitution on spin splitting and magnon behavior. We uncover anisotropic magnon band splitting aligned with electronic structure, and wavevector- and chirality-dependent damping due to Stoner excitations. We identify regions in the Brillouin zone where the chiral magnon splitting overcomes the damping. These findings suggest altermagnets are promising for spintronic and magnonic technologies, where  direction-dependent magnon lifetimes and  nonreciprocal magno transport may enable chiral magnon propagation, while wavevector-selective damping could be harnessed for fast and controllable magnetization switching.

\end{abstract}

\maketitle

\section{\label{sec:level1}Introduction}

Altermagnets represent a newly identified class of magnets that uniquely 
combine characteristics of ferromagnets (FMs) and conventional collinear antiferromagnets (AFMs). For example, consistent with AFMs, altermagnets feature a compensated magnetic structure with zero net magnetic moment.
Similar to FMs, they exhibit spin splitting in their electronic bands along specific 
crystallographic directions \cite{vsmejkal2020crystal, yuan2021prediction,mazin2021prediction,
vsmejkal2022beyond}.  Unlike conventional AFM, this spin splitting 
does not arise from spin-orbit coupling (SOC), instead, it emerges from the interplay between 
magnetic exchange interactions and crystal symmetry. This interplay leads to anisotropic 
spin polarization within the Brillouin zone (BZ), breaking time-reversal symmetry while preserving 
crystal symmetries such as rotations, i.e., $E^{\uparrow}(\textbf{k}) \neq E^{\downarrow}(-\textbf{k})$ without SOC \cite{vsmejkal2022beyond}. The collinear nature of altermagnets implies that the spin remains a good quantum number in the absence of SOC and in difference to non-collinear antiferromagnets, the spin–momentum locking features a common k-independent quantization axis across the Brillouin zone denoting the spin-splitting just as a spin-up, -down splitting. The quantization axis can be changed, by rotating the antiferrmagnetic N\'eel vector relative to the cyrstal lattice with implications on the anomalous Hall effect (AHE), the Dzyaloshinskii-Moriya interaction (DMI) and the current-induced exchange-coupling torques.
These characteristics suggest
that altermagnets represent a distinct category of materials that deviate from conventional
classifications of magnetic order and exhibit unconventional physical properties.

A key aspect of understanding the distinctive behavior of altermagnets lies in the study of 
their spin excitations, which determine their dynamic and transport properties. Spin excitations 
in magnetic systems encompass both collective magnon (spin-wave) modes and single-particle spin-flip 
Stoner excitations \cite{blundell2001magnetism, moriya2012spin}. In FMs, magnons exhibit a quadratic 
dispersion near the Brillouin zone center, resulting in low-energy excitations. The ferromagnetic 
resonance frequency, corresponding to zero-momentum spin excitations in an external magnetic field, 
typically lies in the gigahertz (GHz) range, depending on material parameters and the applied 
field \cite{kittel1951theory}. In contrast, AFMs feature magnons with linear dispersion and significantly 
higher antiferromagnetic resonance frequencies, which are in the terahertz (THz) regime, due to the additional contribution of the strong exchange 
interactions \cite{keffer1952theory}. Altermagnets bridge these two regimes, combining FM-like spin 
splitting with AFM-like symmetry and frequency response, while also exhibiting unconventional magnonic 
properties, including the emergence of chiral magnons, which distinguish them from conventional 
magnetic systems in the absence of SOC \cite{costa2024giant,liu2022switching}.

The term chiral magnons has recently gained attention \cite{vsmejkal2023chiral,chen2024magnons}, yet its 
precise meaning and implications often remain unclear. The concept of magnon polarization 
has been known for decades, and describes the handedness of spin-wave (magnon) precession in magnetic systems. 
According to classical  (the Landau-Lifshitz equation) and quantum mechanics,  a magnetic moment arizing from electrons precesses 
counterclockwise around an applied magnetic field, which is conventionally defined as having positive polarization. 
In simple FMs, all magnons share uniquely this positive polarization \cite{cooke1973neutron}, whereas in collinear 
AFMs, two magnon branches with opposite polarization exist but remain degenerate unless easy-axis or easy-plane anisotropies or a large magnetic 
field lift the degeneracy \cite{keffer1952theory,ross2015antiferromagentic}. Despite these differences, 
direct experimental verification of opposite magnon polarization remains challenging, often requiring polarized 
neutron scattering techniques \cite{maleyev1995investigation,Garst_2017,nambu:2020}.

In altermagnets, the chiral magnon degeneracy of antiferromagnets is lifted along certain wave-vector directions and chiral magnons 
emerge as a consequence of exchange interactions and crystal symmetry, rather than from spin-orbit coupling. 
Unlike chiral magnets, where chirality originates from the Dzyaloshinskii-Moriya interaction in noncentrosymmetric 
systems, altermagnetic chiral magnons arise from the momentum-dependent spin splitting enforced by symmetry, 
even in centrosymmetric materials \cite{vsmejkal2023chiral}. This leads to nonreciprocal magnon dispersions, 
breaking time-reversal symmetry in a distinct way compared to conventional magnetic systems. The resulting 
asymmetric magnon propagation in altermagnets bears similarities to chiral magnets but stems from a fundamentally
different microscopic origin \cite{costa2024giant}. Experimentally, the detection of chiral magnons in 
altermagnets has been demonstrated using inelastic neutron scattering, which has successfully been employed 
to observe altermagnetic magnon splitting in materials such as MnTe \cite{liu2024chiral}. 

Beyond their fundamental significance, chiral magnons in metallic altermagnets exhibit additional 
complexities due to their coupling with Stoner excitations, which significantly affect both their dispersion 
and damping. This coupling leads to wavevector-dependent magnon broadening, distinguishing metallic 
altermagnets from their insulating counterparts and enabling a high degree of control over magnon 
propagation~\cite{costa2024giant, paischer2024electronic,suresh2021magnon}. While such damping may limit 
long-range magnon coherence, it can be functionally advantageous for other applications, such as ultrafast 
magnetization switching, where enhanced damping facilitates rapid relaxation~\cite{wang2019magnetization,choi2025magnetization}. 
These characteristics establish altermagnets as a promising platform for next-generation spintronic and 
magnonic devices, where direction-dependent magnon lifetimes enable nonreciprocal 
transport~\cite{sourounis2024efficient, neusser2009magnonics, chumak2015magnon, baltz2018antiferromagnetic, vsmejkal2022emerging,yu2025neel} 
and symmetry-controlled damping could be harnessed for efficient, field-free switching.

Despite these promising characteristics, the theoretical description of chiral magnons in metallic 
altermagnets remains challenging due to their interaction with Stoner excitations. In particular, 
the classical Heisenberg model, which successfully captures magnon spectra in insulating altermagnets 
\cite{vsmejkal2023chiral,sandratskii2025direct}, fails to describe the coupling between collective 
magnons and single-particle Stoner excitations in metallic systems. Furthermore, it does not account 
for electronic correlation effects, which critically influence magnon lifetimes and damping mechanisms. 
A more accurate framework is provided by many-body perturbation theory (MBPT)~\cite{aryasetiawan1999green,karlsson2000spin}, which explicitly incorporates 
dynamic electron-hole interactions and the full complexity of the electronic structure. This approach has been  applied to the study of elementary ferromagnets and compounds, offering insights into spin  dynamics beyond the limitations of classical models \cite{karlsson2000spin,kotani2008spin,lounis2010dynamical}. We have realized  the MBPT in the context of the full-potential linearized augmented plane wave (FLAPW) methodology in terms of the implementation of the \textsc{SPEX} code~\cite{csacsiouglu2010wannier,friedrich18,SPEX2}. The code has been applied to a number of ferromagnets~\cite{csacsiouglu2010wannier,friedrich2014spin} and served as benchmark for an equivalent TDDFT implementation of the magnetic response function~\cite{Skovhus:2021}.

In this paper, we employ MBPT combined with density functional theory (DFT) to investigate chiral magnon excitations in a family of metallic 3$d$ transition-metal monopnictides. The materials considered—VSb, CrSb, CrAs, and CrBi—exhibit  A-type antiferromagnetic N\'eel order and crystallize in the centrosymmetric hexagonal NiAs-type structure, where the transition metal \(T =\) V, Cr and  the pnictogen  \(Pn =\) As, Sb, Bi. These compounds exemplify \textit{g}-wave altermagnetism, wherein the momentum-dependent spin splitting of electronic bands transforms under spatial symmetries according to an octupolar harmonic ($\ell=4$) with even parity, resembling a cloverleaf pattern with eight nodes.  We systematically 
analyze the relationship between electronic band spin splitting and magnon dispersions, with a 
particular focus on how pnictogen substitution influences these characteristics.

\begin{figure}[!t]
\includegraphics[width=\columnwidth]{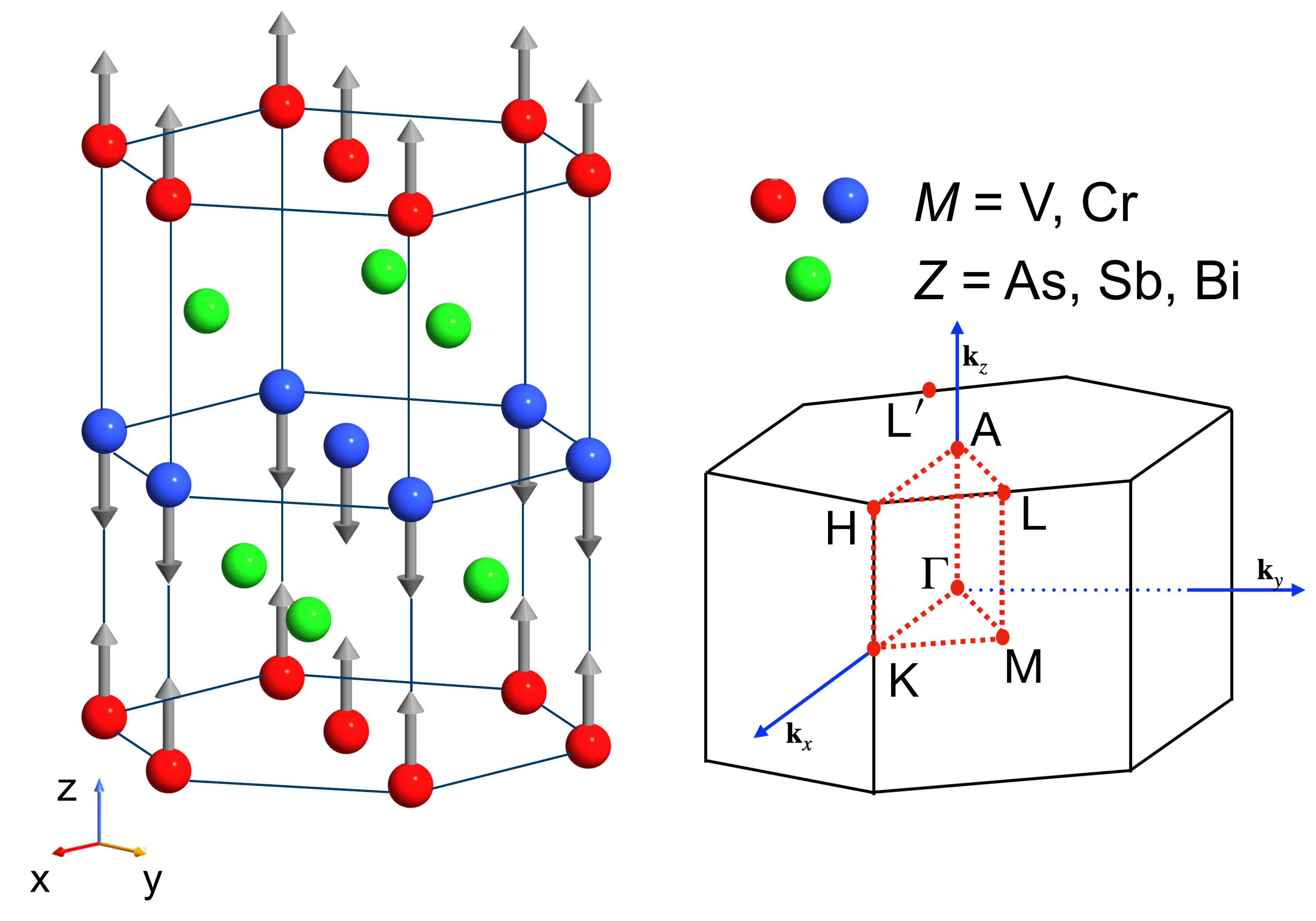}
\vspace{-0.65cm}
\caption{\label{fig1} 
Left: Schematic representation of the magnetic and crystalline structure of the NiAs-type altermagnetic \( TPn \) compounds  (\( T = \) V, Cr; \( Pn = \) As, Sb, Bi), illustrating the compensated A-type layered antiferromagnetic ordering of  ferromagnetic sublattices (blue and red transition metal atoms) with opposite magnetic moments (arrows). $z$ is parallel to the antiferromagnetic translation direction $[0001]$.
Right: Brillouin zone of the hexagonal lattice, with high-symmetry points indicated.}
\end{figure}

The selection of these materials is motivated by several factors. One is certainly the NiAs-type crystal structure, illustrated in Fig.~\ref{fig1}. This structure features alternating ferromagnetic atomic layers of transition metals stacked antiferromagnetically along the $z$ axis separated by pnictogen atoms in a hexagonal arrangement. The $T$-$Pn$ superexchange along the $z$-axis stabilizes the  magnetic configuration with antiparallel sublattice moments, resulting in a compensated 
net-zero magnetic moment system. The triangular arrangements of the pnictogen atoms   above and below each  transition-metal  sublattice are rotated by $60^\circ$ with respect to each other, which is the structural framework that supports altermagnetic behavior. The high crystal symmetry (\(P6_3/mmc\)) prevents conventional linear DMI between neighboring atoms and by this weak ferromagnetism. Thus, the magnetic moments are really collinear with a uniaxial N\'eel vector along the $z$-direction.

Experimentally, VSb and CrSb are well-established in this phase 
\cite{grison1962crystal,kjekshus1969properties,liu2024chiral}, while CrAs can crystallize  
either in the NiAs-type  or MnP-type (orthorhombic, \(Pnma\)) structure, depending 
on synthesis conditions \cite{selte1973phase}.  For CrBi, no direct experimental confirmation of its stable structure exists, though theoretical studies propose the NiAs-type as its ground state, with metastable alternatives like the zincblende (ZB cubic \(F\bar{4}3m\) structure considered in spintronic contexts \cite{reisi2017study}. Other factors motivating the transition-metal monopnictides are their relatively simple binary composition,
which facilitates synthesis with fewer structural and chemical defects, as well as the relatively low-energy glide planes of the NiAs structure that makes altermagnetic properties susceptible  to strain,  making them attractive for spintronic applications. 

 However, the first and most important point for the selection of the material family was that among the investigated compounds CrSb emerges as a benchmark system of altermagnets due to its high N\'{e}el  temperature $T_{\mathrm{N}}$ ($>700$ K) and significant electronic spin splitting ($>1$ eV) observed experimentally near the Fermi level~\cite{zeng2024observation, ding2024large, Reimers:2024, yang2025three}. In addition, recent experimental work on the magnon modes of CrSb~\cite{biniskos2025systematic} has provided direct 
evidence of chiral magnon excitations and their energy splitting. 
This establishes CrSb as a key reference material for metallic  
altermagnetic spin dynamics and highlights the necessity of further investigations into related 
compounds to explore the broader implications of chiral magnon excitations. Besides studying the details of the prototypical CrSb system, it is important to study a set of similar materials, in order to gain understanding and intuition of the remarkable altermagnetic properties combining spin-splitting and chiral magnon splitting in a compensated magnetic structure through the study of chemical trends.

Our results reveal a pronounced anisotropic magnon band splitting that closely mirrors the 
momentum-dependent spin splitting in the electronic bands. Furthermore, we demonstrate that 
magnon damping due to Stoner excitations exhibits significant wavevector dependence, with 
strong broadening in specific regions of the Brillouin zone while remaining moderate in others. 
This variation highlights the intricate coupling between magnon excitations and the underlying 
electronic structure of metallic altermagnets. Beyond providing new insights into magnon behavior 
in metallic systems, our findings establish altermagnets as tunable platforms for directional 
magnon transport and control. These characteristics position altermagnets as promising candidates 
for next-generation spintronic and magnonic devices, where tailored magnon lifetimes and 
nonreciprocal transport could enable energy-efficient signal processing and high-frequency 
applications.

\section*{\label{sec:level2}Results} 
\subsection*{\label{sec:sub1_level2}Electronic and magnetic properties} 
To explore the unique properties of metallic altermagnets, we investigate the structural, electronic, and 
magnetic characteristics of $TPn$ compounds ($T =$ V, Cr; $Pn =$ As, Sb, Bi) using DFT (for details see Method Sections). The results are collected in Table~\ref{table1}.
Denoted by the negative sign of $E_{\mathrm{AFM-FM}}$, our calculations confirm that the collinear layered $c$-axis AFM configuration, \text{i.e.}\ magnetic structure with an A-type N\'eel vector with an easy axis parallel to [0001] as shown in Fig.~\ref{fig1}, has lower energy than FM. The energy difference $E_{\mathrm{AFM-FM}}$ gives a rough estimate of the N\'eel temperatures. Estimating the N\'eel temperature in the mean field approximation (MFA) as $k_\text{B}T_\text{N}=\frac{2}{3} E_{\mathrm{AFM-FM}}$ and taking into account that the MFA overestimates Monte Carlo results by about 59\% for layered systems~\cite{Jia:2020}, we estimate the N\'eel temperature of CrSb to 875~K, which is in the ballpark of the experimental value of 700~K. The resulting magnetic anisotropy energy (MAE) exhibits an out-of-plane preference for all $TPn$ compounds as indicated by the negative sign of $E_{\mathrm{MAE}}$ in Table~\ref{table1}. We therefore assume that this is the magnetic ground state. To systematically determine the equilibrium lattice parameters, 
we performed structural optimizations for both FM and AFM states. The relaxed lattice constants for 
each configuration were found to be different, indicating that magnetic ordering influences the 
structural parameters. However, since the AFM state was consistently lower in energy across all 
materials, we adopted the AFM lattice parameters as the reference values in Table~\ref{table1} and 
used them for comparing the total energy differences between the FM and AFM states. This choice 
ensures a consistent basis for energy comparisons, eliminating any discrepancies arising from 
structural relaxation effects. The observed preference for the AFM ground state highlights the 
dominance of strong  magnetic exchange interactions that favor compensated magnetic ordering, 
which is a characteristic feature of altermagnetic systems.

\begin{table*}
\caption{\label{table1}  
Lattice parameters ($a$ and $c$), total energy difference per formula unit between antiferromagnetic (AFM) and 
ferromagnetic (FM) calculations, $E_{\mathrm{AFM-FM}}$, total energy difference  of AFM order between  out-of-hexagonal plane and in-plane orientation of N\'eel vector, $E_{\mathrm{MAE}}$, spin and orbital magnetic  moments, $m_{T}$, $m_{\mathrm{orb},T}$, 
maximum spin splitting, $\Delta E^{\uparrow\downarrow}$, and maximum magnon band splitting, 
$\Delta \omega^{\uparrow\downarrow}$, in metallic g-wave altermagnets $TPn$ ($T=$ V, Cr; 
$Pn=$ As, Sb, Bi). The experimental lattice parameters are available for CrSb and VSb in Ref.~\cite{yang2025three,lomnytska2007phase,Reimers:2024}. Top block: optimized minimum energy  lattice parameters. Bottom block: lattice parameters fixed to the values of CrSb.} 
\begin{ruledtabular}
\begin{tabular*}{\textwidth}{@{\extracolsep{\fill}}lccclcccc}
$\mathrm{TPn}$ & $a$ & $c$ & $E_{\mathrm{AFM-FM}}$ & $E_{\mathrm{MAE}}$ &$m_{T}$ & $m_{\mathrm{orb},T}$ &$\Delta E^{\uparrow\downarrow}$ & $\Delta \omega^{\uparrow\downarrow}$ \\
              & (\AA) & (\AA) & (meV) & (meV) &($\mu_\text{B}$) &  ($\mu_\text{B}$) & (eV) & (meV) \\ \hline
VSb  & 4.10 & 5.79 & \phantom{0}$-$12& $-$0.008 & 1.77  & 0.000 & 0.55 & 40 \\ 
CrAs & 3.77 & 5.35 & \phantom{0}$-$95 & $-$0.09 &2.51 & 0.007 & 1.18  &18 \\
CrSb & 4.11 & 5.44 & $-$180 & $-$0.42 &2.79 & 0.015 & 1.25 & 52 \\  
CrBi & 4.40 & 5.52 & $-$280 & $-$0.67 &3.24 & 0.059 & 1.19 & 36 \\ 
\hline
VSb  & 4.11 & 5.44 & \phantom{00}$-$5 & $-$0.008 &1.57 & 0.000 & 0.43 & 39 \\ 
CrAs & 4.11 & 5.44 & $-$200 & $-$0.20 & 3.03 & 0.010 & 1.10 & 25 \\
CrSb & 4.11 & 5.44 & $-$180 & $-$0.42 & 2.79 & 0.015 & 1.25 & 52 \\ 
CrBi & 4.11 & 5.44 & $-$170 & $-$1.15 & 2.85 & 0.072 & 1.35 & 43 \\ 
\end{tabular*}
\end{ruledtabular}
\end{table*}

\begin{figure*}[t]
\includegraphics[width=2.07\columnwidth]{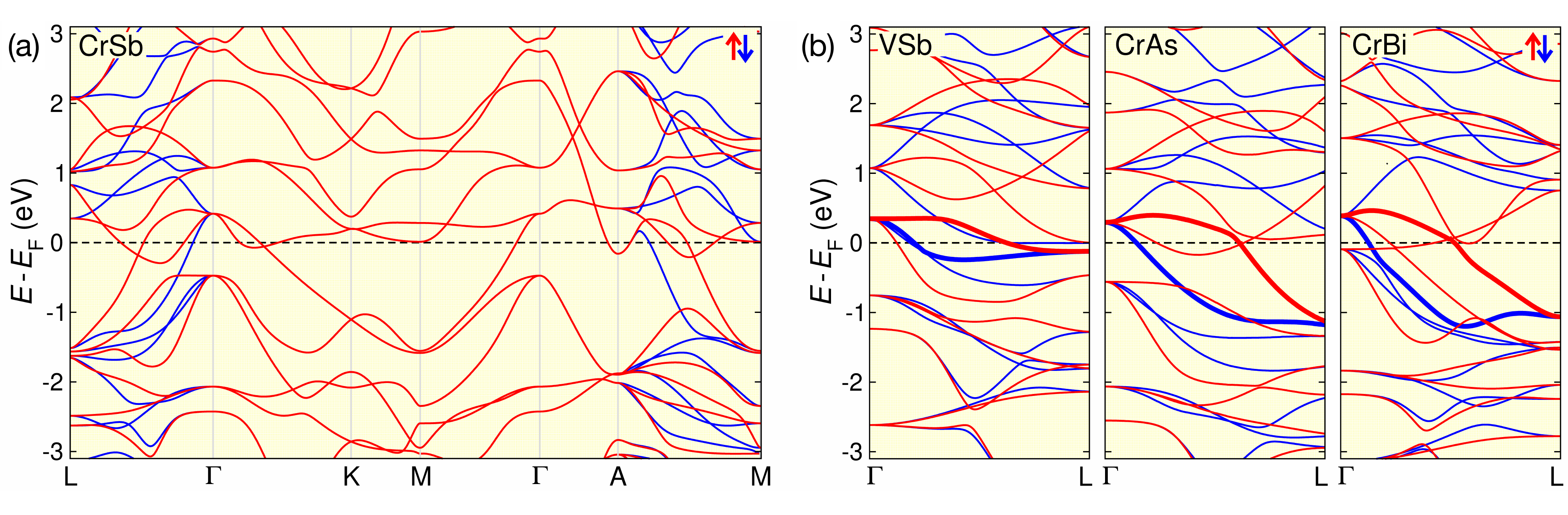}
\vspace{-0.75 cm}
\caption{\label{fig2}
(a) Spin-resolved band structure of CrSb along 
the path L-$\Gamma$-K-M-$\Gamma$-A-M in the Brillouin zone. 
(b) Spin-resolved band structures of VSb, CrAs, and CrBi along the $\Gamma$-L 
direction, highlighting regions with significant spin splitting. Spin-split 
bands near the Fermi level are emphasized for clarity.  The dashed lines denote the  
Fermi level, which is set to zero energy.}
\end{figure*}

 Table~\ref{table1} summarizes the lattice parameters ($a$ and $c$), sublattice
magnetic moments ($m_{\mathrm{V/Cr}}$), and maximum spin splitting values ($\Delta E^{\uparrow\downarrow}$) along the path $\Gamma$-L
for all compounds. CrSb exhibits a large sublattice magnetic moment ($2.79~\mu_\text{B}$) and a substantial 
spin splitting of 1.25~eV in its relaxed lattice configuration, reflecting strong magnetic exchange interactions. 
In contrast, VSb shows a smaller magnetic moment ($1.77~\mu_\text{B}$) and reduced spin splitting of 0.55~eV, 
demonstrating the critical role of the transition metal atom in governing the magnetic and electronic properties. 
These results highlight how the choice of transition metals and pnictogen elements directly influences the 
fundamental characteristics of altermagnets, impacting both their electronic structure and magnetic behavior.

To disentangle the effects of lattice geometry and compositional variation, we performed 
calculations for all compounds with lattice parameters fixed to those of CrSb, as shown in 
the second block of Table~\ref{table1}. This approach allows us to isolate the impact of 
chemical composition from structural effects, providing deeper insight into the intrinsic 
electronic and magnetic interactions in these materials. Under this constraint, the spin 
splitting, $\Delta E^{\uparrow\downarrow}$,  of VSb decreases to 0.43~eV, while CrBi 
increases to 1.35~eV. These trends reveal 
that the electronic structure is highly sensitive to lattice variations, with larger unit 
cells tending to enhance spin splitting. The pronounced dependence of spin splitting on 
lattice parameters demonstrates their pivotal role in modulating the electronic and magnetic 
interactions in altermagnets, further demonstrating how both structural and compositional 
factors collectively shape their fundamental properties.

The spin-resolved band structure of CrSb, presented in Fig.~\ref{fig2}(a), reveals significant 
spin splitting near the Fermi level, particularly along the $\Gamma$-L and A-M directions in the 
Brillouin zone. Given the previously 
discussed sensitivity of spin splitting to lattice parameters, we now examine the role of chemical 
composition by comparing the electronic band structures of VSb, CrAs, and CrBi along the same
direction, as shown in Fig.~\ref{fig2}(b) for relaxed lattice parameters. This comparison highlights 
how pnictogen substitution  modifies the electronic structure, with CrSb exhibiting the largest 
spin splitting among the studied compounds, followed by CrBi under the relaxed lattice conditions. 
The observed trend aligns with the  previously established correlation between larger lattice 
parameters and enhanced spin splitting, further reinforcing the interplay between structural and 
compositional effects in altermagnets.

\subsection*{\label{sec:sub2_level2}Spin excitations}

The electronic and magnetic properties of $TPn$ compounds significantly influence their spin excitation 
spectra. Magnons, which are collective spin-wave excitations in magnetic materials, are described by 
the transverse magnetic response function $R^{+-}(\mathbf{q}, \omega)$, see methods section for details. Given the strong spin splitting observed 
in the electronic band structures, it is crucial to analyze how these electronic properties translate 
into magnon dynamics. Within MBPT, the transverse magnetic response function is schematically given by: 
$R^{+-}=K^{+-}+K^{+-}T^{\downarrow\uparrow}K^{+-}$, where the first term represents the response of the noninteracting 
system (the Kohn-Sham magnetic response function), and the second term describes interaction effects via the 
$T$ matrix. The $T$ matrix accounts for repeated scattering events of particle-hole pairs with opposite spins, 
which ultimately leads to the formation of collective magnon excitations. It is schematically expressed as: 
$T^{\downarrow\uparrow}=[1-WK^{+-}]^{-1}W$, where $W$ is the screened Coulomb interaction. The magnetic response function 
thus provides a framework for understanding both collective magnon modes and single-particle spin-flip Stoner 
excitations, along with their respective lifetimes.

In ferromagnets, magnons reduce the total magnetization by one quantum of angular momentum, $\hbar$, 
and are associated with the spin-lowering operator $S^-$. This process corresponds to an electron transition 
from the spin-up ($\uparrow$) band to the spin-down ($\downarrow$) band. 
The attractive electron-hole interaction $W$ leads to the formation of a bound state—the magnon—whose energy 
depends on the wave vector $\mathbf{q}$. This bound state manifests as a well-defined peak in $R^{+-}(\mathbf{q}, \omega)$, 
representing the collective magnon mode, while single-particle spin-flip Stoner excitations appear as a 
higher-energy continuum. The distinction between magnons and Stoner excitations is fundamental to understanding 
spin dynamics, as the former corresponds to coherent magnons, whereas the latter leads to damping effects 
that limit magnon lifetimes.

In AFMs, magnons exist in two forms: those with \( S_z = -1 \), which lower the spin in the spin-up (\(\uparrow\))
sublattice, and those with \( S_z = +1 \), which raise the spin in the spin-down (\(\downarrow\)) sublattice~\cite{costa2024giant}. Due to the 
symmetry between the two sublattices, these magnon modes are degenerate across the Brillouin zone. However, altermagnets 
exhibit a fundamentally different behavior: their transverse magnetic response functions, \( R^{+-}(\mathbf{q}, \omega) \) 
and \( R^{-+}(\mathbf{q}, \omega) \), become non-degenerate along specific crystallographic directions where electronic 
spin splitting occurs. This non-degeneracy leads to chiral magnon excitations, giving rise to two distinct magnon branches, 
denoted as \( \omega^- \) and \( \omega^+ \) bands. The resulting anisotropic magnon transport differs from conventional 
AFMs, where magnon propagation is symmetric. In directions without spin splitting, the two response functions 
remain degenerate, leading to a time-reversal symmetric magnon spectrum.

\begin{table}[!t]
\caption{\label{table2} 
Average screened on-site direct $W$/$W'$ (diagonal/off-diagonal) and exchange $J$ 
Coulomb matrix elements between the 3$d$ orbitals of the $T$ atom and the $p$ orbitals 
of the $Pn$ atom in $TPn$ compounds ($T=$ V, Cr; $Pn=$ As, Sb, Bi). Values are provided 
for both their computationally optimized lattice parameters and for lattice parameters fixed to 
those of CrSb.}
\begin{ruledtabular}
\begin{tabular}{llcccc} 
&  &   \multicolumn{2}{c}{[$a$,$c$]$_{\mathrm{th}}$} &   \multicolumn{2}{c}{$[a,c]_{\mathrm{CrSb}}$} \\ \cline{3-4} \cline{5-6} 
 $TPn$ & Orbital  & $W$/$W'$ (eV) & $J$ (eV)  & $W$/$W'$ (eV) & $J$ (eV)\\ 
\hline
VSb  & V-$3d$   &  0.80/0.19   & 0.31 & 0.82/0.20  & 0.31 \\ 
     & Sb-$4p$  &  1.17/0.65   & 0.26 &  1.17/0.65  & 0.26  \\
CrAs & Cr-$3d$  &  1.36/0.59   & 0.38 &  1.54/0.69  & 0.43 \\
     & As-$3p$  &  1.48/0.85   & 0.32 &  1.37/0.78  & 0.10  \\
CrSb & Cr-$3d$  &  1.51/0.69   & 0.41 &  1.51/0.69   & 0.41\\ 
     & Sb-$4p$  &  1.25/0.72   & 0.26 &  1.25/0.72  & 0.26  \\    
CrBi & Cr-$3d$  &  1.51/0.66   & 0.43 &  1.45/0.59  & 0.43 \\ 
     & Bi-$5p$  &  1.18/0.67   & 0.26 &  1.30/0.75  & 0.28 \\
\end{tabular}
\end{ruledtabular}
\end{table}

\begin{figure*}[tp]
\includegraphics[width=2.07\columnwidth]{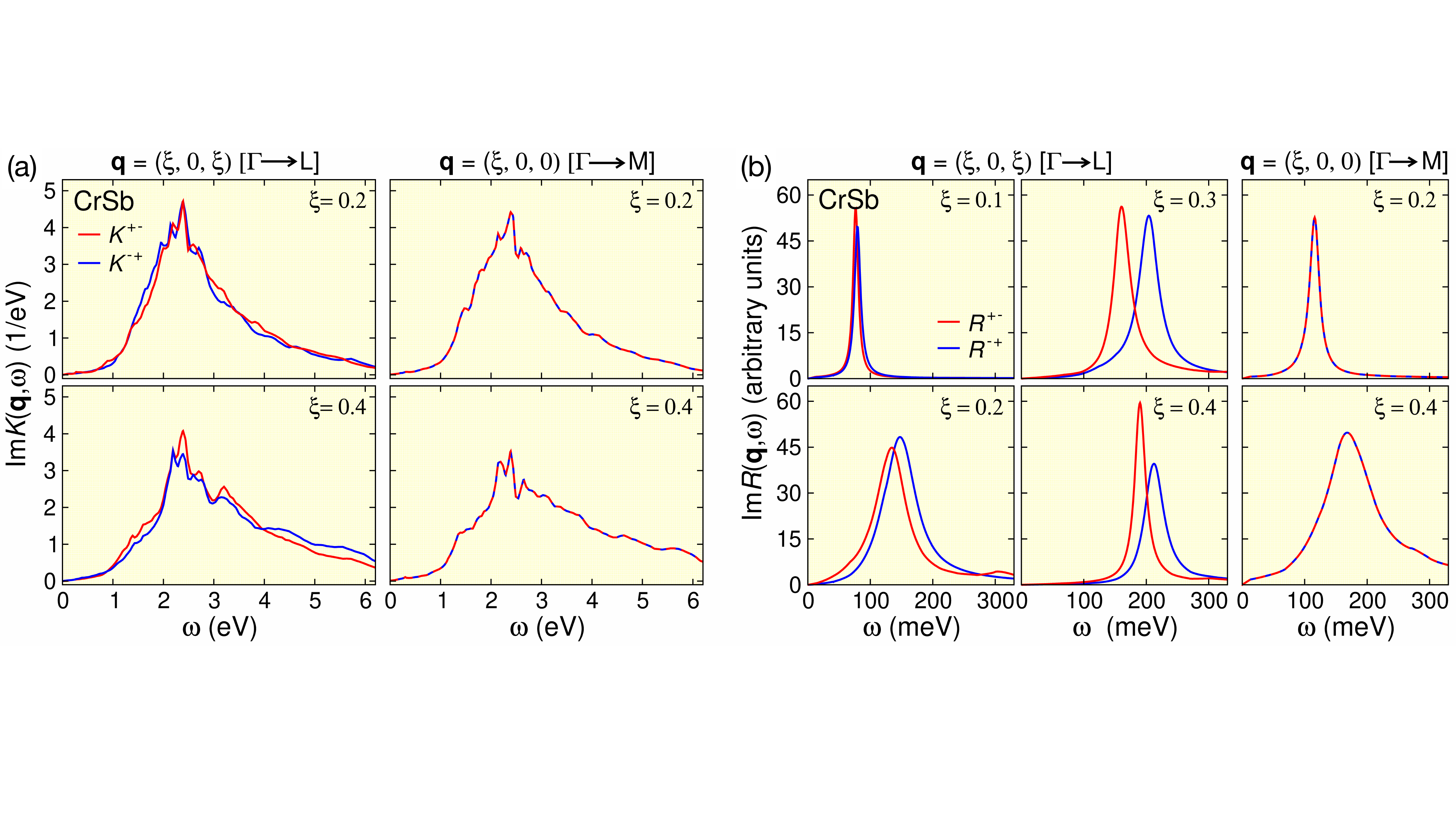}
\vspace{-2.6 cm} 
\caption{\label{fig3} 
(a)  Noninteracting Kohn-Sham magnetic response functions \(K^{+-}\) (red line) and \(K^{-+}\) (blue line) 
for CrSb, plotted for selected wave vectors along the \(\Gamma\)-L and \(\Gamma\)-M directions in the 
Brillouin zone. Since electron-electron interactions are absent at this level, the spectral function 
exhibits only single-particle Stoner excitations, with no  collective magnon modes at lower energies.
The broad continuum of excitations corresponds to spin-flip transitions between occupied and 
unoccupied electronic states of opposite spin. The non-degeneracy of \(K^{+-}(\mathbf{q},\omega)\) and 
\(K^{-+}(\mathbf{q},\omega)\) along \(\Gamma\)-L reflects the underlying spin splitting in the electronic 
band structure, while their degeneracy along \(\Gamma\)-M suggests a lack of intrinsic spin asymmetry 
in this direction.  
(b) Magnon excitation spectra for CrSb, obtained from the response functions \(R^{+-}(\mathbf{q},\omega)\) 
(red line) and \(R^{-+}(\mathbf{q},\omega)\) (blue line) for selected wave vectors along the \(\Gamma\)-L and 
\(\Gamma\)-M directions. Compared to the noninteracting case, additional well-defined magnon peaks emerge 
at lower energies due to electron-electron interactions. The response functions remain non-degenerate along 
\(\Gamma\)-L, giving rise to chiral magnon excitations, while they remain  degenerate along \(\Gamma\)-M. 
For clarity, peak amplitudes in each panel are scaled to the same height for comparative visualization. The wave vector $\mathbf{q}=(q_1,q_2, q_3)= q_1\mathbf{b}_1 + q_2\mathbf{b}_2 +q_3\mathbf{b}_3$ is given in units of the reciprocal lattices vectors $\mathbf{b}_1, \mathbf{b}_2, \mathbf{b}_3$. $\Gamma:(0,0,0)$, L: $(1/2,0,1/2)$, M: $(1/2,0,0)$. For more details see Ref.~\cite{SM}.}
\end{figure*}  

To establish a quantitative foundation for $R^{+-}(\mathbf{q}, \omega)$, we first start from the AFM state and compute the matrix elements 
of the onsite screened Coulomb potential $W$ for the 3$d$ orbitals of the $T$ atom and the $p$ orbitals of 
the $Pn$ atom in $TPn$ compounds ($T=$ V, Cr; $Pn=$ As, Sb, Bi). These values, presented in Table~\ref{table2}, 
are calculated for both computationally optimized lattice parameters and those fixed to CrSb. Given that the
strength of electron-electron interactions plays a crucial role in determining magnon excitations, understanding 
the trends in $W$, $W'$ and $J$ provides key insights into the interplay between Coulomb screening and spin dynamics 
in these materials.  The screened Coulomb interaction $W$ (and $W'$) for the 3$d$ orbitals of V and Cr is comparable 
to their elementary bulk bcc metal counterparts, indicating moderate electron correlation effects. For 
pnictogen $p$ orbitals, the screened Coulomb interaction decreases from As to Sb to Bi, following 
the expected trend of increasing atomic size and polarizability. This variation influences the 
hybridization between the transition metal $d$ states and pnictogen $p$ states, affecting the electronic 
and magnetic properties of the compounds. When the lattice parameters are fixed to those of CrSb, the 
$W$ values show slight changes across all compounds. The exchange interaction $J$ remains relatively stable 
across compounds, with values ranging from 0.26 to 0.43~eV. CrBi  exhibits the highest $J$ among Cr-based 
compounds, reflecting stronger localization of Cr-3$d$ orbitals, while VSb displays lower $J$, consistent 
with its smaller exchange splitting and enhanced screening effects.

\begin{figure*}[tp]
\includegraphics[width=2.05\columnwidth]{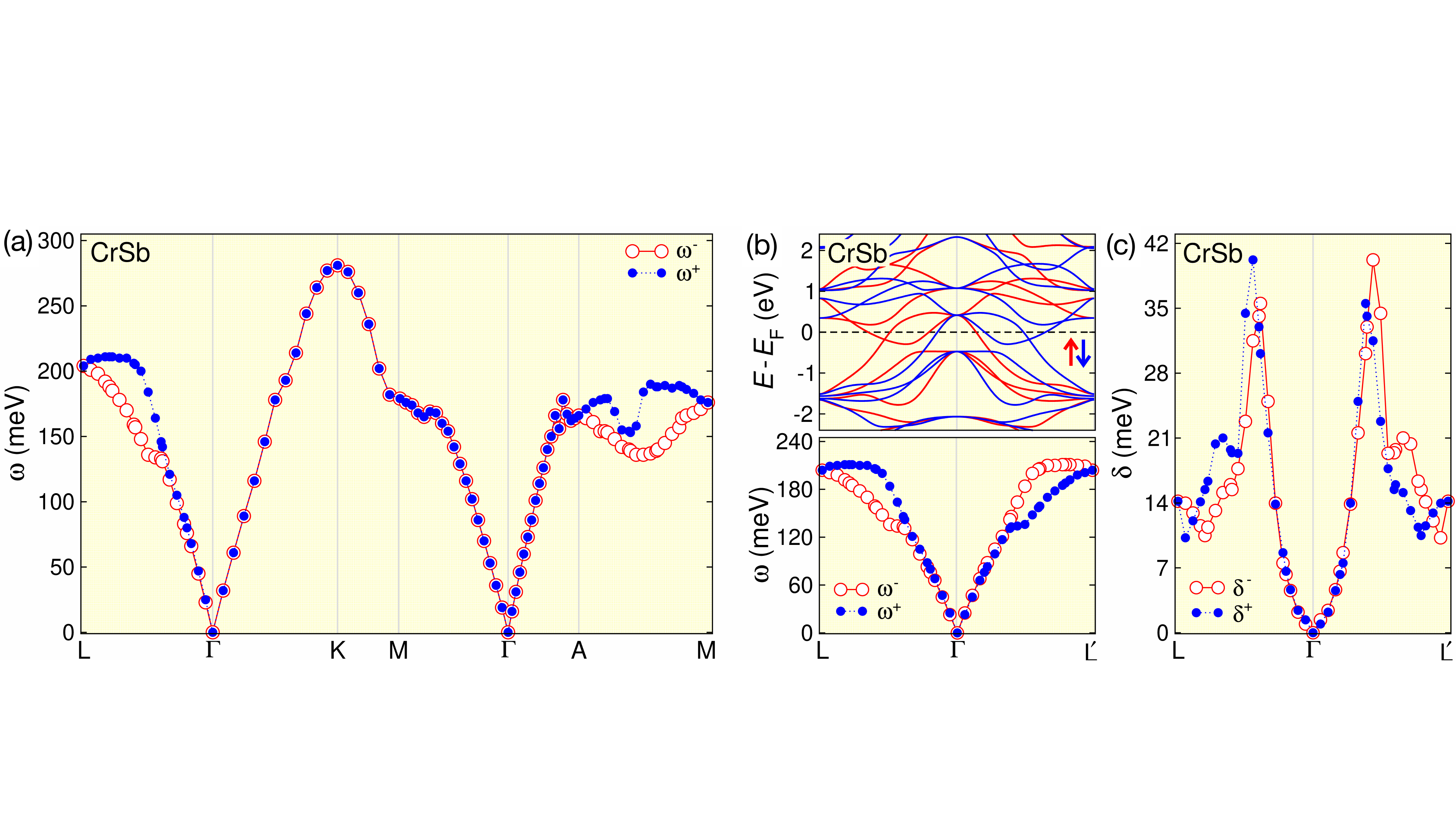}
\vspace{-1.8 cm}
\caption{\label{fig4} 
(a) Magnon dispersion for CrSb along the L-$\Gamma$-K-M-$\Gamma$-A-M path in the Brillouin zone, 
showing two distinct magnon branches, $\omega^-$ (red) and $\omega^+$ (blue). 
(b) Upper panel: Spin-resolved electronic band structure for CrSb along the L-$\Gamma$-L$'$ path, 
highlighting spin-up (red) and spin-down (blue) bands with significant spin splitting. 
Lower panel: Magnon dispersion along the same path, displaying both $\omega^-$ and $\omega^+$ branches. 
(c) Magnon lifetimes along the L-$\Gamma$-L$'$ path, represented by the half-width at half-maximum (HWHM) 
of the magnon peaks for both branches. These results illustrate the connection between electronic spin splitting 
and magnon properties, including dispersion and damping.}
\end{figure*}

Stoner excitations, which play a key role in understanding magnon damping, arise from electron transitions 
between bands of opposite spin. When an electron is excited from an occupied majority-spin state at 
$\mathbf{k}$ to an unoccupied minority-spin state at $\mathbf{k} + \mathbf{q}$, it generates an electron-hole 
pair with a triplet spin configuration, reducing the magnetization by unity. Since these are single-particle 
spin-flip processes, they can be qualitatively studied at the Kohn-Sham level via the imaginary part of the 
non-interacting magnetic response function, $\textrm{Im} K(\mathbf{q}, \omega)$.  In Fig.~\ref{fig3}(a), we 
present $\textrm{Im} K(\mathbf{q}, \omega)$ for CrSb at selected wave vectors along the $\Gamma$-$\textrm{L}$ and 
$\Gamma$-$\textrm{M}$ directions. Since renormalization effects due to electron-electron interactions are 
absent at this level, only single-particle spin-flip Stoner excitations are observed. The spectral function, 
$\textrm{Im}K^{+-}(\mathbf{q}, \omega)$ [$\textrm{Im}K^{-+}(\mathbf{q}, \omega)$], exhibits a broad peak 
due to spin-flip transitions between occupied majority and unoccupied minority states. The peak maximum, 
located around 2.5~eV, corresponds to the exchange splitting of the Cr sublattice in CrSb, as confirmed by 
the projected density of states in the Supplementary Information~\cite{SM}.  As illustrated in Fig.~\ref{fig3}(a), the 
non-interacting magnetic response functions $\textrm{Im}K^{+-}(\mathbf{q}, 
\omega)$ and $\textrm{Im}K^{-+}(\mathbf{q}, \omega)$ exhibit distinct behavior depending on the electronic 
spin splitting. Along the $\Gamma$-L direction, where significant spin splitting occurs in the band structure, 
the two response functions are non-degenerate. However, along the $\Gamma$-M direction, where electronic spin 
splitting is absent, the response functions remain degenerate. This highlights the intrinsic connection between 
the spin-dependent electronic structure and single-particle spin-flip excitations in altermagnets.

When dynamical correlations are included via the screened Coulomb interaction \( W \) 
(see Table~\ref{table2}), additional low-energy magnon peaks emerge in the spectral functions 
of the interacting system, \(\text{Im} R^{+-}(\mathbf{q}, \omega)\) and \(\text{Im} R^{-+}(\mathbf{q}, 
\omega)\), as shown in Fig.~\ref{fig3}(b) 
for CrSb. Similar to the non-interacting response function, the interacting magnetic response 
functions, \( R^{+-}(\mathbf{q}, \omega) \) and \( R^{-+}(\mathbf{q}, \omega) \), are 
non-degenerate along the \(\Gamma\)-L direction due to the intrinsic spin splitting in the 
electronic band structure. For small wave vectors \(\mathbf{q}\), the magnon peaks are nearly 
degenerate and exhibit sharp spectral features. As \(\mathbf{q}\) increases, the peaks split 
and broaden, indicating variations in magnon lifetimes. This broadening arises from magnon 
coupling to single-particle Stoner excitations, which provide additional decay channels. 
In contrast, along the \(\Gamma\)-M direction, where spin splitting is absent, both response 
functions remain degenerate, reflecting the symmetry of the underlying electronic structure.

Due to the metallic nature of CrSb, Stoner excitations exist at all energies, leading to substantial 
magnon damping for finite wave vectors (\(\mathbf{q} \neq 0\)). Unlike in insulating magnets, where 
magnons are well-defined collective modes with long lifetimes, in metallic systems, they can decay via 
coupling to the continuum of single-particle spin-flip Stoner excitations. This decay mechanism is particularly
relevant in CrSb, where the overlap between the magnon and Stoner excitation spectra results in a significant
broadening of the magnon peaks in both \(\text{Im} R^{+-}(\mathbf{q}, \omega)\) and \(\text{Im} 
R^{-+}(\mathbf{q}, \omega)\), as illustrated in Fig.~\ref{fig3}(b). The degree of magnon damping is 
directly linked to the density of states (DOS) of Stoner excitations and the electronic DOS in the 
corresponding energy range (see Supplementary Information~\cite{SM}). A higher DOS of Stoner excitations 
at a given energy increases the probability of magnon decay, as more decay channels become available. 
This effect is particularly pronounced for large \(\mathbf{q}\), where the magnon dispersion enters 
energy regions with a high Stoner DOS, resulting in stronger damping. Conversely, for small \(\mathbf{q}\), 
where the magnon energy remains low, the overlap with Stoner excitations is reduced, leading to sharper 
and more well-defined magnon peaks. These trends highlight the strong correlation between the electronic 
structure and magnon lifetimes in altermagnets, demonstrating how the interplay between spin excitations 
and electron-hole continuum governs the stability of collective magnon modes in metallic systems.

\begin{figure*}[ht!]
\includegraphics[width=1.66\columnwidth]{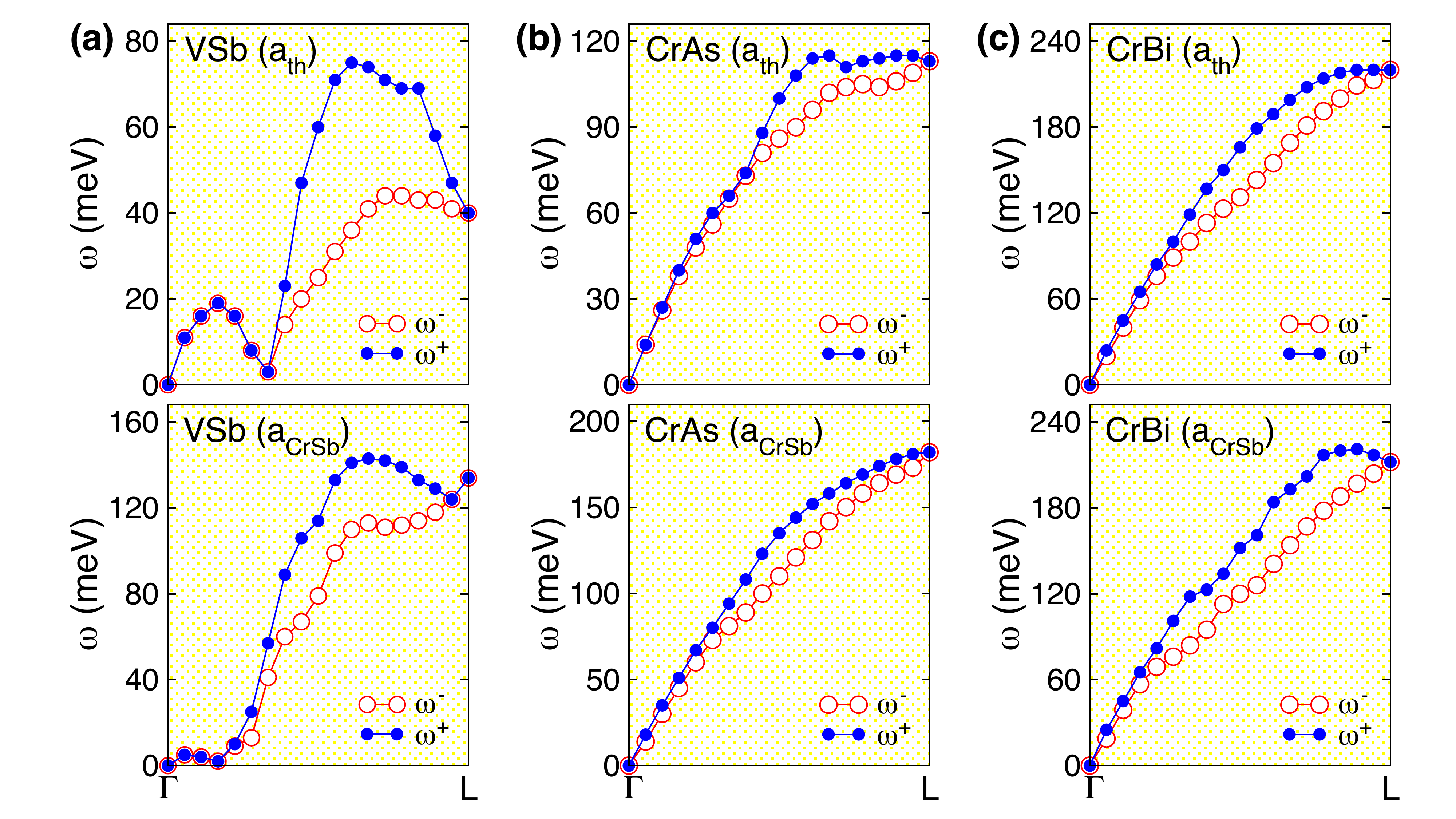}
\vspace{-0.3 cm}
\caption{\label{fig5} 
Magnon dispersion along the \(\Gamma\)-L direction for (a) VSb, (b) CrAs, and (c) CrBi, 
showing two distinct magnon branches, \(\omega^-\) (red) and \(\omega^+\) (blue). 
The upper row presents dispersions calculated using the theoretical lattice parameters of each material, 
while the lower row displays results obtained with the lattice parameters fixed to those of CrSb. Notice, the specific energy scale of each panel.}
\end{figure*}

Since the magnon energies are determined from the transverse magnetic response functions, 
\( \text{Im} R^{+-}(\mathbf{q}, \omega) \) and \( \text{Im} R^{-+}(\mathbf{q}, \omega) \), 
they correspond to the positions of the spectral peaks in these functions. By plotting these 
peaks as a function of the wave vector \( \mathbf{q} \), we obtain the magnon dispersion, as 
shown in Fig.~\ref{fig4}(a) along the L-\(\Gamma\)-K-M-\(\Gamma\)-A-M path in the Brillouin 
zone for CrSb. As expected, a linear magnon dispersion, $E=Aq$, is observed near the \(\Gamma\) point, 
consistent with the behavior of gapless Goldstone modes in metallic magnets. The slope of the dispersion $E=Aq$ gives a spin stiffness $A=201$ meV \AA. 
Similar to the 
electronic band structure, we find magnon band splitting along the \(\Gamma\)-L and A-M directions, 
reflecting the intrinsic spin splitting characteristic of altermagnetic materials. The two distinct 
magnon branches, denoted as \( \omega^- \) and \( \omega^+ \), arise due to this splitting and 
correspond to the energy positions of the spectral peaks in \( \text{Im} R^{+-}(\mathbf{q}, \omega) \) 
and \( \text{Im} R^{-+}(\mathbf{q}, \omega) \), respectively. The maximum splitting reaches 
approximately 52~meV, highlighting a substantial energy difference between the two magnon branches 
in these directions. This splitting is a direct consequence of the nontrivial spin-dependent 
interactions in CrSb and further demonstrates the anisotropic nature of its magnon spectrum \cite{vsmejkal2023chiral}. 
The maximum magnon energy is found at the K point, reaching up to 285~meV, which is significantly 
higher than the maximum energies at L (200~meV) and M (175~meV). This variation in magnon energies 
across the Brillouin zone reflects the wavevector-dependent exchange interactions in CrSb. The 
lower energy at M compared to K suggests that the effective spin-exchange interactions are weaker 
along the corresponding crystallographic directions, leading to a reduced magnon group velocity. These 
findings emphasize the intricate connection between band structure and magnon dispersion, 
demonstrating how electronic spin splitting directly manifests in the magnon spectrum of metallic 
altermagnets.

Figure~\ref{fig4}(b) illustrates the chiral nature of the magnon dispersion in CrSb by comparing the 
electronic band structure (upper panel) and magnon spectrum (lower panel) along the L-\(\Gamma\)-L\('\) 
directions. A striking feature of the magnon dispersion is the interchange of split magnon branches 
between the \(\Gamma\)-L and \(\Gamma\)-L\('\)  paths, indicating a lack of inversion symmetry in magnon 
propagation. This directional dependence of magnon energies is a hallmark of chiral magnons, where 
the dispersion relation varies depending on the propagation direction in momentum space.  
Unlike conventional antiferromagnets, where magnon modes remain degenerate in opposite directions, 
CrSb exhibits a pronounced spin asymmetry in its magnon bands. This asymmetry reflects the intrinsic chiral 
character of altermagnets, arising from momentum-dependent spin splitting in the electronic structure. 
As a result, CrSb possesses anisotropic magnon transport, which could give rise to nonreciprocal spin-wave 
dynamics and unconventional magnonic effects. 

To gain insight into magnon lifetimes, we analyze the half-width at half-maximum (HWHM) of 
the interacting spectral function along the L-\(\Gamma\)-L\('\) direction for CrSb, as shown 
in Fig.~\ref{fig4}(c). The magnon lifetimes along the  full high-symmetry path L-\(\Gamma\)-K-M-\(\Gamma\)-A-M in the Brillouin can be found in Fig.~S3 of the Supplementary Information~\cite{SM}. In our notation, the HWHM, which corresponds to the magnon damping rate, 
is denoted by \(\delta\) in the figure. The magnon lifetime \(\tau\) is then given by the 
relation \(\tau = \hbar/\delta\), where \(\hbar = 6.582 \times 10^{-4}\) eV·ps is 
the reduced Planck’s constant. Our results show that among the split magnon branches, the 
higher-energy mode consistently exhibits larger HWHM values, indicating stronger damping for 
magnons with higher excitation energies. This trend is expected, as higher-energy magnons 
have a larger phase space for decay into Stoner excitations, leading to shorter lifetimes. 
Along the \(\Gamma\)-L and \(\Gamma\)-L\('\)  directions, the HWHM follows a quadratic increase from the \(\Gamma\) point reaching a peak value of 40~meV at an intermediate wave vector before decreasing non-monotonically with oscillations to about 10 meV at the L (L\('\) )
point. These oscillations suggest an interaction between magnons and the underlying 
electron-hole excitations in different momentum regions. This damping can be approximated as $\delta = \alpha q^2$, with $\alpha=85$ meV \AA$^{2}$ for $q\in\{\Gamma, \text{L}\}$ and $55$ meV \AA$^{2}$ for $q\in\{\Gamma, \text{K}\}$. This directional dependence of $\alpha$ highlights the anisotropy in the damping behavior along different directions in momentum space.  The highest magnon damping occurs along 
the \(\Gamma\)-M direction (see Fig.~S3 in the Supplementary Information~\cite{SM}), where the HWHM 
reaches 60 meV near the M point, corresponding to the shortest magnon lifetime. In contrast, 
along the \(\Gamma\)-K path, the damping remains 
relatively smooth, with a maximum HWHM of 20 meV, indicating longer-lived magnons in this direction. The magnon lifetime at the A point at the end of the $\Gamma$-A line along $k_z$ direction yields 40~meV. 
These findings demonstrate the strong wavevector dependence of magnon lifetimes in CrSb and 
highlight how the interplay between magnon excitations and Stoner decay channels governs 
magnon stability in metallic altermagnets. From the estimated values of HWHM, we find that magnon 
lifetimes in CrSb vary from a few femtoseconds in strongly damped regions (e.g., near the
M point, see Supplementary Information~\cite{SM}) to over 60~fs in regions of lower damping (e.g., at the L point). 
This variation highlights the critical role of wavevector-dependent Stoner excitations in determining 
the stability of magnons. The relatively smooth HWHM profile along the \(\Gamma\)-K direction suggests 
that this path is more favorable for long-lived magnon transport, whereas the pronounced damping near 
M indicates efficient magnon decay into electron-hole pairs.

To extend our analysis beyond CrSb, we present the magnon dispersions along the $\Gamma$-L direction for 
VSb, CrAs, and CrBi in both their relaxed lattice parameters and in the CrSb lattice constant, as 
summarized in Table~\ref{table1} and shown in Fig.~\ref{fig5}. These results reveal substantial variations 
in magnon energies and band splitting across different compounds, reflecting the interplay between lattice 
structure, magnetic moments, and electronic spin splitting. Among the three compounds, VSb exhibits the 
most striking behavior, featuring a strong magnon band splitting of 40~meV in its relaxed structure, 
comparable to CrSb. However, its maximum magnon energy remains relatively low, reaching only 75~meV along 
the $\Gamma$-L direction. When the lattice parameters are fixed to those 
of CrSb, the magnon bands change substantially; while the maximum magnon band splitting remains nearly the same, 
the magnon energies nearly double, highlighting a strong lattice-dependent modification of magnon dynamics. For CrAs, 
the magnon dispersion follows an expected linear increase near the $\Gamma$ point before reaching 120~meV in its relaxed
structure and 170~meV in the CrSb lattice. This increase suggests that CrAs possesses higher-energy magnons than 
VSb, consistent with its larger magnetic moment (2.51~$\mu_\text{B}$ vs.\ 1.77~$\mu_\text{B}$ in VSb). Despite this, the magnon band 
splitting remains relatively modest at 18~meV in the relaxed lattice and 25~meV in the CrSb lattice, indicating that 
its chiral magnon effects are less pronounced compared to CrSb and VSb. In contrast, CrBi exhibits high-energy magnons, 
reaching 210~meV in both lattice settings, with a moderate magnon band splitting of 36~meV in the relaxed lattice and 
43~meV in the CrSb lattice. While CrBi has a strong exchange interaction, as reflected in its high magnetic moment 
(3.24~$\mu_\text{B}$), its chiral magnon effects are weaker than in CrSb. However, at the L point, CrBi and CrSb display 
similar magnon energies, indicating that both compounds have comparable effective exchange couplings in this momentum 
region. This suggests that despite differences in overall magnon band splitting and anisotropy, CrSb and CrBi share 
fundamental similarities in their exchange interactions at specific points in the Brillouin zone.

\section*{\label{sec:level3}Discussion}

Our DFT results are in good agreement with known experimental and computation results. Concerning the experimental data, we mostly compare with CrSb, as this system is well-studied theoretically and experimentally. The relaxed lattice parameters of CrSb lie within the ball park of earlier studies~\cite{yang2025three,Reimers:2024}. The lattice parameters  of VSb deviate approximately  by 4\% and 6\% from the experimental $a$ and $c$ values, respectively~\cite{lomnytska2007phase}. To the best of our knowledge, there are no experimental studies reporting CrAs or CrBi in the NiAs-type crystal structure. The calculated magnetic easy axis along the $z$-direction for CrSb is in line with prior results~\cite{park2020effects}, and the spin splitting near the Fermi level agrees well with experimental observations~\cite{yang2025three}. The energetics of our calculations are in the right ball park of experimental results. For example, we roughly estimated the Curie temperature of CrSb to 875~K instead of the experimental value 700~K and the  antiferromagnetic resonance energy, discussed below, to 12~meV in excellent agreement with the 12.3~meV measured at 11~K.

Key characteristic of the magnon dispersion of altermagnets is the chiral magnon splitting. Our MBPT study demonstrates that metallic \( g \)-wave altermagnets (\( TPn \), where \( T = \mathrm{V}, \mathrm{Cr} \);
\( Pn = \mathrm{As}, \mathrm{Sb}, \mathrm{Bi} \)) with A-type N\'eel vector exhibit a typical antiferromagnetic magnon dispersion with an additional chiral split magnon excitations, in the same reciprocal space directions where electronic bands show spin splitting. Among the investigated compounds, CrSb shows the strongest magnon band  splitting of 52~meV at about $\mathbf{q}=(0.28,0,0.28)$ (in units of the reciprocal lattice vectors), i.e., somewhat beyond middle of the $\Gamma$-L path. A similar large splitting we find on the path A-M around $(0, 0, 0.25)$, but with rapid variations along the path. However, we did not find a one-to-one correspondence between the size of the magnetic moment, the electronic spin-splitting and the chiral magnon splitting. For example, Cr in CrBi in the bulk lattice constant has the largest Cr magnetic moment, but the largest spin-splitting and the largest chiral magnon splitting is exhibited by CrSb. On the other hand,  CrBi strained to the lattice constant of CrSb, has the largest spin-splitting but not the largest chiral spitting of the magnon excitation. CrSb shows only the second highest magnon energy of about 200~meV along the path $\Gamma$–L although the magnon-splitting is largest. The chiral splitting of CrSb has recently been determined also by linear spin wave theory on top of Heisenberg exchange parameters determined by DFT and a maximum value of 22~meV was found on the path  $\Gamma$–L~\cite{biniskos2025systematic}, which is significantly smaller than our MBPT data.

VSb, on the other hand, combines a moderate 
magnon band splitting with relatively low-energy excitations. These low-energy magnons fall within the typical 
detection range of inelastic neutron scattering (INS), suggesting that VSb could be a viable candidate for future 
experimental studies of chiral split magnons. Interestingly,  VSb displays on an otherwise smooth and monotonically increasing trend a sudden and pronounced dip followed by a sharp recovery in the magnon dispersion at an intermediate wavevector along the \( \Gamma \)–L direction indicative of a magnetic phase transition  toward a more complex 
spin-spiral ground state. Taking VSb in the lattice constant of CrSb, the sudden drop off in the dispersion is absent, which implies that this phase transition  is susceptible to strain and alloying.  This warrants further investigation beyond the scope of this work.

An important topic is magnon damping, which can occur through the decay of the magnon via the electron-electron interaction into the Stoner continuum, through the spin-orbit interaction and through magnon-magnon scattering. In this work, the focus is on the former. Since the systems are metallic, this relaxation channel is expected to provide the largest contribution to damping. Damping shows a very interesting behavior across the BZ. It can be different for the two chiral modes. Of course, we expect that the damping increases with the magnon energy. In CrSb, damping is particularly strong along the \( \Gamma \)-M direction (80~meV, see Fig.~S3 in Supplementary Materials~\cite{SM}), it is about 15~meV at the L point, about 20~meV at the K point. It remains  relatively weak around the \( \Gamma \) point leading to long-lived magnons in specific regions of the Brillouin zone. From the $\Gamma$ point the damping  grows quadratically with the wave vector into the BZ. 

Most interesting is the competition between magnon splitting and magnon damping. Therefore, we inspect the two splitting active symmetry lines $\Gamma$-L and A-M with particular attention.  
For CrSb, we find a maxiumx attenuation of the magnon mode  of 40~meV at about $\mathbf{q}=(0.21,0,0.21)$ (more precisely, we find two different wave vectors of maximum attenuation depending on the chirality,  $\mathbf{q}_+=(0.22,0,0.22)$ and $\mathbf{q}_-=(0.19,0,0.19)$, which are close to the $\mathbf{q}$ point of largest spin splitting (see above). However, the magnon damping varies rapidly in $\mathbf{q}$ space, and, by $\mathbf{q}=(0.28,0,0.28)$, the magnon damping has strongly decreased to $\delta_+=19$ and $\delta_-=17$~meV. Furthermore, away from $\mathbf{q}=(0.28,0,0.28)$ towards the L point, the spin splitting remains sizable, while the damping is weak. We conclude that the magnon spin splitting is best observed experimentally around the center of  the $\Gamma$-L path, somewhat closer to the L point. 
The observability of the splitting can also be read off from Fig.~\ref{fig3}(b), where the experimentally measurable response functions develop between  $ (\tfrac{1}{5}, 0, \tfrac{1}{5})$ and $ (\tfrac{2}{5}, 0, \tfrac{2}{5})$ two energy-split peaks of two different intensities for the two magnons of opposite chirality. The maximum peak splitting of 52~meV and the two line widths at HWHM of $\delta_++\delta_-=36$~meV sets a measure for the experimentally required energy resolution of about 20-30~meV at 180~meV excitation energy. Concerning the A-M direction, for the magnon damping we find again a strong wave vector, but also a strong chirality dependence (for details see Fig.~3 in Ref.~\cite{SM}). At the wave vector of maximum magnon splitting (49 meV), the damping $\delta_+=25$ meV and $\delta_-=20$ meV are about the same size as the splitting energy. Both findings show that the experimental observability of the chiral magnon splitting can be very wave vector sensitive. 
The damping ratios ($\delta/\omega$), or the inverse of the quality factors of the magnon mode, respectively, vary along the path and reach values of $\delta/\omega=0.07$ at the long-life L point and 0.2 for the short-life mode half-way on the high-symmetry path between $\Gamma$ and L. The ratios are comparable to typical values reported for other metallic magnets~\cite{friedrich2014spin}. Since single crystals for CrSb are grown as plates where the c-axis is perpendicular to the plate, inevitably this is always like a reference direction for the growth of the crystal. Therefore, we would like to encourage the measurement of the magnon lifetime for the A point (see BZ in Fig.~\ref{fig1}), for which we found a value of 40~meV (HWHM).

 To benchmark our results for CrSb, we compare the extracted damping parameters, $\alpha_{\Gamma\text{-K}} \approx 55$ meV$\cdot$\AA$^{2}$, and $\alpha_{\Gamma\text{-L}} \approx 85$ meV$\cdot$\AA$^{2}$, and spin-stiffness parameters, $A_{\Gamma\text{-K}}=200$ meV$\cdot$\AA, and $A_{\Gamma\text{-L}}=175$ meV$\cdot$\AA\ with inelastic neutron scattering measurements reported by Radhakrishna and Cable~\cite{Radhakrishna:1996}. Their analysis revealed anisotropy in CrSb, with stiffness constant of $A=266$ meV$\cdot$\AA, and damping coefficient $\alpha = 110$ meV$\cdot$\AA$^{2}$ along $\Gamma$-K direction. Thus, our values are in reasonable agreement with the experimental values of $\alpha$ and $A$, but on the smaller side. Within the linear spin wave theory approximation, as reported in \cite{biniskos2025systematic}, the spin stiffness is approximately 280 meV \AA. The discrepancies between our MBPT results and experiment may be attributed to the higher measurement temperature in the experiment (295 K). Additionally, our model does not account for  SOC effects and  magnon-magnon scattering, which is expected to contribute in the experimental conditions.

A recent experimental study using resonant inelastic X-ray scattering (RIXS) with circular polarization has provided 
the first direct evidence for the existence of chiral magnon excitations in CrSb~\cite{biniskos2025systematic}.  The observed circular dichroism in the RIXS 
spectra supports the presence of two oppositely polarized magnon branches, which is consistent with our theoretical 
findings. However, the limited energy resolution of the experiment prevented a direct observation of the magnon band 
splitting, and its effects were inferred only through polarization-dependent intensity variations. The study measures 
magnon energies along two distinct directions in the Brillouin zone: along the \(\Gamma\)-A direction, where no 
altermagnetic magnon splitting is expected, multiple \(\mathbf{q}\)-points were probed, allowing for a more complete 
mapping of the dispersion. In contrast, along the M–A direction, where altermagnetic splitting of magnon bands is 
predicted, only a single \(\mathbf{q}\) point was measured, restricting a more detailed comparison with the theoretical 
dispersion. This limited momentum-space coverage prevents a direct experimental confirmation of the full altermagnetic 
magnon band structure predicted by our MBPT calculations.

Despite the experimental constraints limiting a direct verification of magnon splitting, they  allow for a meaningful 
comparison of overall magnon energies between RIXS measurements and theoretical models. The RIXS measurements 
indicate that the magnon dispersion reaches approximately 140~meV at the A point along the \( \Gamma \)-A direction and 
about 150~meV at the single measured $\mathbf{q}$ point along the A-M direction.  Our MBPT calculations closely match 
the experimental energy at this $\mathbf{q}$ point, while predicting slightly higher energies (162 meV) at the A point. It is interesting to understand how good is actually the comparison between our MBPT approach and an approach based on DFT-extracted exchange parameters combined with a linear spin-wave model. The latter has been worked out recently for CrSb in Ref.~\cite{biniskos2025systematic}. The latter significantly 
underestimates the RIXS magnon energies along \( \Gamma \)-A, yielding only 100 meV at the A point unless an \textit{ad hoc} 
downward shift of the Fermi level by 34 meV is introduced. While our MBPT dispersion is relatively flat along A-M, 
the linear spin-wave calculations predict a significant variation from 100 meV to 215 meV along this path. This suggests that 
itinerant electron effects captured in MBPT~\cite{csacsiouglu2010wannier}, which are absent in linear spin-wave approach, 
may play an essential role in shaping the magnon dispersion. Another key difference lies in the predicted magnon splitting 
along the A-M direction. Our MBPT calculations yield a splitting of 52~meV, whereas the linear spin-wave model predicts only  22 meV. Additional experiments are at need to shed more light computational description of magnon in metallic altermagnets.

The experimental validation of the predicted magnon splitting is  a challenging experimental objective as requirements to resolve it are  at the cutting edge of RIXS, INS and spin-polarized electron energy loss spectroscopy (SPEELS)~\cite{zhang2007magnons,dos2021spin,holbein2023spin,abraham1989spin,vollmer2003spin,qin2017temperature}. All probes are excellent and probe either the response function \( R^{+-}(\mathbf{q}, \omega) \), which governs magnon excitations or allow direct measurements of magnon dispersions and lifetimes by detecting the dynamic structure factor 
\( S(\mathbf{q}, \omega) \), proportional to the imaginary part of \( R^{+-}(\mathbf{q}, \omega) \), but different requirements on the sample size relative to the  available monodomain size and the required energy resolution relative to the probe energy may favor one of the methods over the other or require specific further developments.  We encourage experimental science to address this challenge in the near future. 

The directional 
and momentum-selective damping in metallic altermagnets may be advantageous for ultrafast magnetization switching applications.
In contrast to magnonic logic circuits that require long magnon lifetimes, switching processes benefit from enhanced damping,
which suppresses precessional oscillations and accelerates relaxation. Recent experimental studies have shown that spin angular 
momentum carried by magnons can induce efficient magnetization switching~\cite{wang2019magnetization}, and that magnonic 
spin dissipation can enhance spin--orbit torque--driven reversal in ferromagnetic heterostructures~\cite{choi2025magnetization}. 
In this context, the chiral magnon modes and anisotropic damping inherent to altermagnets could be harnessed to generate 
directional magnonic torques for fast and controllable spin switching. These findings point to a broader functional scope 
for metallic altermagnets in symmetry-enabled, field-free spintronic memory technologies. For energy-efficient magnonic logic 
devices based on coherent magnon transport, very low damping is required, lower than that offered by altermagnetic metals. Here, altermagnetic insulators are alternatives to be explored.

This directional variation in magnon propagation,
combined with wavevector-dependent damping, suggests that altermagnets offer a versatile platform for engineering 
nonreciprocal magnon excitations.
At the end, we briefly comment on the role of the spin-orbit coupling (SOC), that was mostly neglected in this work, but has multiple ramifications such as on the magnon damping, a further splitting on the electronic bands and the chiral magnon energies, the formation of the easy N\`eel axis and together with the exchange interaction it determines the  antiferromagnetic resonance frequency. Since the magnetic moments are in the order of 2 to 3~$\mu_\text{B}$, corresponding to a spin 1 or 3/2 system, combination of exchange and spin-orbit interaction maybe lead to a higher-order biquadratic DMI, not considered sofar. Our calculations with SOC show that for all systems the easy N\'eel vector axis is along the $z$ axis of the NiAs structure, consistent with experimental data for CrSb~\cite{park2020effects}. The anisotropy energy between the easy and hard axis, $E_\text{MAE}$, (see Table~\ref{table1}) increases  with the increase of the  nuclear number of the pnictogen atom. We find small associated orbital moments of Cr whose values increase accordingly. Obviously, due to the $d$-$p$ hybridization between $T$-$Pn$, $Pn$ determines the size of the MAE. With SOC, the band structure depends on the direction of the N\'eel vector relative to the crystal lattice. The electronic band structure calculated with SOC and the  N\'eel vector oriented along the easy $z$ axis of the NiAs lattice is presented for CrSb in Figure~S1 of the Supplementary Information~\cite{SM}. We find that the spin-splitting due to SOC is a few percent of the altermagnetic exchange splitting $\Delta E^{\uparrow\downarrow}$  (in the order of 10-70~meV depending on $\mathbf{k}$ point and band). 
In comparison to the altermagnetic band structure without SOC shown in Fig.~\ref{fig2}(a), the effect of SOC becomes evident in two main aspects. First, SOC lifts band degeneracies along high-symmetry lines that are not altermagnetically active and thus were not previously split. Second, it causes the opening of previously crossing bands within the already split bands along the $\Gamma$–L and M–A directions. Changing the N\'eel vector axis from out-of-plane to in-plane (x-axis) (not shown in Ref.~\cite{SM}) 
does not change the overall SOC picture, but  modifies details of the SOC induced splitting.
From the changes of the band structure in the 10-70~meV range, we conclude that the effect on the chiral magnon splitting is negligible. 

With inclusion of the SOC, an excitation gap $\hbar\omega_\text{AR}$ emerges in the magnon-spectrum of antiferromagnets at $\Gamma$, whose associated frequency  known as antiferromagnetic resonance (AR), is related to the uniform precession modes of the sublattice magnetizations. It is a result of the spin-orbit coupling through the magnetic anisotropy energy, $E_\text{MAE}$, combined with the Heisenberg-type exchange energy, $E_\text{X}$. For easy-axis antiferromagnets with magnetic moments of the two sublattices pointing  in the direction $\pm z$, in absence of an external magnetic field, the gap energy $\hbar\omega_\text{AR} =\sqrt{E_\text{MAE}(E_\text{MAE}+2E_\text{X})}$~\cite{keffer1952theory} of the two modes of the two sublattices are degenerate. Since the $E_\text{MAE}$ is orders of magnitudes smaller than  $E_\text{X}$, we can approximate $\hbar\omega_\text{AR}$ by $\hbar\omega_\text{AR} \simeq \sqrt{2E_\text{MAE}E_\text{X}}$. For antiferromagnets, the magnetic anisotropy energy is simply the magnetocrystalline energy, \textit{i.e.}\ the energy difference between the easy and hard N\'eel vector axis. Focusing on CrSb, the exchange energy per atoms can be read off as $E_\text{X}=E_{\mathrm{AFM-FM}}$ from Table~\ref{table1} to $E_\text{X}=180$~meV and  $E_\text{MAE}=0.42$~meV, which results into a resonance energy of 12.3~meV in excellent agreement to the measured value of 12~meV at 11~K \cite{Radhakrishna:1996}.

\section*{\label{sec:level3}Methods}

Our study employs a combination of two different DFT methods and MBPT to describe the structural, electronic, 
magnetic, and spin excitations in the considered metallic \textit{g}-wave altermagnets (\(TPn\), where \(M =\) V, Cr;
\(Z =\) As, Sb, Bi).  The computational workflow comprises three main steps: (i) structural optimization using the 
\textsc{QuantumATK} code~\cite{QuantumATK,QuantumATKb}, (ii) electronic and magnetic structure calculations using the 
full-potential linearized augmented-plane-wave (FLAPW) method implemented in the \textsc{FLEUR} code~\cite{fleurCode}, 
and (iii) spin excitation calculations using the T-matrix approach, which is implemented within the \textsc{SPEX} 
code~\cite{Spex,SPEX2,csacsiouglu2010wannier} and utilizes Wannier functions generated by the \textsc{Wannier90} code~\cite{mostofi2008wannier90,pizzi2020wannier90}.

\subsection*{\label{sec:sub1_level3}Crystal structure optimization}

The equilibrium lattice parameters (\textit{a, c}) of all studied compounds were determined using the
\textsc{QuantumATK} software package \cite{QuantumATK,QuantumATKb}. We used a linear combination of atomic 
orbitals (LCAO) as a basis set, together with norm-conserving PseudoDojo pseudopotentials~\cite{VanSetten2018} 
and the Perdew-Burke-Ernzerhof (PBE) parametrization of the GGA functional \cite{Perdew1996}. The Brillouin 
zone was sampled using a $16 \times 16 \times 12$ Monkhorst-Pack $\mathbf{k}$-point grid \cite{Monkhorst1976}. 
Structural relaxation was carried out until the forces on each atom were reduced below \(10^{-4}\) eV/Å, 
ensuring accurate determination of the lattice constants.
\subsection*{\label{sec:sub2_level3}Electronic and magnetic structure calculations}

Following structural optimization, electronic and magnetic property calculations were conducted using the 
full-potential linearized augmented plane wave (FLAPW) method, as implemented in the \textsc{FLEUR} code \cite{fleurCode}. 
The GGA functional with PBE parameterization \cite{Perdew1996} was used to describe exchange-correlation effects. The basis set, augmented for V and Cr atoms by local 3s and 3p orbitals, is included up to a  reciprocal cutoff radius of \(K_\mathrm{max} = 4\) bohr\(^{-1}\) and an angular momentum cutoff
of \(l_\mathrm{max} = 10\) within the muffin-tin spheres. The muffin-tin sphere radii were set to 2.51 a$_0$ and 2.59 a$_0$ for Cr and Sb in CrSb, respectively. In the relaxed structures, CrAs had radii of 2.35 a$_0$ for both Cr and As, while CrBi had 2.54 a$_0$ for Cr and 2.80 a$_0$ for Bi. Similarly, in VSb, both V and Sb were assigned a radius of 2.56 a$_0$. Compounds adopting the lattice parameters of CrSb retained the same muffin-tin radii as Cr and Sb. The Brillouin zone was sampled with a $16 \times 16 \times 12$ $\mathbf{k}$-point mesh, ensuring well-converged 
electronic structures. Two different magnetic structures had been studied, the FM and the collinear layered $c$-axis AFM order. The calculations where carried out in the same unit-cell to produce reliable energy differences.

SOC effects are in principle neglected in this work and are therefore also initially neglected in the electronic and magnetic structure calculations. To better understand the magnetic ground state and the energy gap at the $\Gamma$-point relevent for the antiferromagnetic resonance, we determined the easy magnetization axis and the crystalline magnetic anisotropy energy $E_\text{MAE}$. We carried out self-consistent total energy calculations  including SOC for two magnetization directions, along the out-of-plane, $z$ direction and along the in-plane, $x$ axis. The energy difference amounts to $E_\text{MAE}$. Convergence was achieved with a 41$\times$41$\times$30 k-point set in the full BZ. This allows us to determine the easy axis and orbital moments of V and Cr for all compounds in their relaxed ground state structures. The impact of SOC in the out-of-plane direction on the band structure of CrSb is presented in Fig.~S1 in the Supplementary Information~\cite{SM}.

\subsection*{\label{sec:sub3_level3}Spin excitation calculations}

The spin-wave calculations are based on MBPT. We outline the main theoretical framework in the following and refer the reader to Refs.~\onlinecite{csacsiouglu2010wannier,friedrich18} for details. The transverse magnetic response function is formally given 
by the functional derivative 
$R^{+-}(1,2) = \delta m^+(1)/\delta B^+(2)$ 
with the  magnetization density $m^+(1)$ and a perturbing magnetic field $B^+(2)$ where $1=(\mathbf{r},t)$, 2, etc.~are space-time coordinates. Here, the superscript "$+$" refers to a right-handed (with respect to the majority spin direction) sense of circular polarization, which identifies $R^{+-}(1,2)$ as the \emph{transverse} spin susceptibility; $\delta m^+$ and $\delta B^+$ lie in the plane perpendicular to the spin polarization. 
The connection to Green-function theory is established by $m^+(1)=-2i G^{\downarrow\uparrow}(1,1^+)$ with the time-ordered single-particle Green function $G^{\sigma\sigma'}(1,2)$ and $1^+=(\mathbf{r},t+0^+)$. 
Carrying out the functional differentiation and using the Dyson equation ${G^{\sigma\sigma'}}^{-1}={G_0^{\sigma\sigma'}}^{-1}-\Sigma^{\sigma\sigma'}$ (in matrix notation) with the non-interacting Green function and the $GW$ approximation for the electronic self-energy $\Sigma^{\sigma\sigma'}(1,2)=iG^{\sigma\sigma'}(1,2)W(1,2)$ leads to
\begin{align}
R^{+-}(1,2)&=iG^\downarrow(1,2)G^\uparrow(2,1)
-\iiiint G^\downarrow(1,3)G^\uparrow(4,1) \nonumber\\
&\times T^{\downarrow\uparrow}(3,4;5,6)G^\downarrow(5,2)G^\uparrow(2,6)
d3\,d4\,d5\,d6~,\label{eq:Rmat}
\end{align}
where the $T^{\downarrow\uparrow}$ matrix fulfills the Bethe-Salpeter equation
\begin{align}\label{eq:Tmat}
T^{\downarrow\uparrow}(1,2;3,4) &= W(1,2)\delta(1,3)\delta(2,4)+i\iint W(1,2)\\
&\times G^\downarrow(1,5)G^\uparrow(6,2)T^{\downarrow\uparrow}(5,6;3,4)
d5\,d6~.\nonumber
\end{align}
The $T$ matrix and also the two-particle Green function $iG^\downarrow(1,2)G^\uparrow(3,4)$ depend on four points in space-time. In order to be able to perform calculations, we have to introduce a few simplifications and approximations. First, the fact that the Hamiltonian is time-independent reduces the number of independent time variables to three. We then assume that $W(1,2)$ acts instantaneously, i.e., the two time variables are identical. This leaves us with a single time variable and, after Fourier transformation, with a single frequency $\omega$. As a function of $\omega$, $W(1,2;\omega)$ is constant, and we set $W(1,2;\omega)=W(1,2;\omega=0)$. This static approximation is justified by the fact that magnon energies lie far below the material's plasma frequency, where $W(\omega)$ varies slowly.

We then represent all quantities, including the two-particle Green function $K^{+-}=iG^\downarrow G^\uparrow$, in terms of a product basis $\{w_{\mathbf{R}n}(\mathbf{r})w_{\mathbf{R}'n'}(\mathbf{r}')\}$ of maximally localized Wannier functions (MLWFs) $\{w_{\mathbf{R}n}(\mathbf{r})\}$ where $\mathbf{R}$ denotes an atom position vector. This allows us to truncate the screened interaction $W_{\mathbf{R}n,\mathbf{R}'n'}=W_{\mathbf{R}n,\mathbf{R}n'}\delta_{\mathbf{R}\mathbf{R}'}$. Inserting the solution of the Bethe-Salpeter equation, Eq.~(\ref{eq:Tmat}), into Eq.~(\ref{eq:Rmat}) and projecting from left and right onto a plane wave $\exp(i\mathbf{kr})$ then finally yields the magnetic response function $R(\mathbf{k},\omega)$.
The $T$-matrix approach sketched above is 
implemented in the \textsc{SPEX} code. The maximally localized Wannier functions (MLWFs) were constructed 
from first-principles FLAPW calculations using the \textsc{Wannier90} library. MLWFs provide an effective 
low-energy representation for the $T$-matrix calculations.  The correlated Hilbert space of the \(TPn\) compounds 
includes the 3$d$ orbitals of Cr (or V) and the $p$ orbitals of the \(Pn\) atom.  Given that the unit cell contains 
four atoms, we constructed a total of 16 Wannier orbitals, comprising 10 orbitals with 3$d$ character and 6 orbitals 
with $p$ character. The screened Coulomb interaction \(W\) was computed within the random-phase approximation (RPA) using 
a mixed product basis set. For this calculation, we included 350 unoccupied states, and the resulting screened interaction 
was subsequently projected onto the MLWF basis.  
The screened Coulomb interaction \(W\) was computed using a 
coarse \(\mathbf{k}\)-point mesh of \(12 \times 12 \times 8\), as it converges rapidly with respect to \(\mathbf{k}\)-point
sampling. In contrast, the evaluation of \(R\) required a denser \(\mathbf{k}\)-point mesh of \(40 \times 40 \times 40\) 
to ensure numerical accuracy.

 In practical implementations, the magnetic response function \( K \) and the screened Coulomb interaction \( W \) are computed using Kohn–Sham Green's functions, which can lead to a violation of the Goldstone condition \cite{mueller2016}. To correct for this and ensure gapless magnon excitations at \( \mathbf{q} \rightarrow 0 \), we apply a material-dependent scaling \( W \rightarrow \zeta W \). This adjustment restores the correct low-energy behavior of the magnon dispersion without altering its overall features. The values of \( \zeta \) used for each material are provided in the Supplementary Material.

\begin{acknowledgments}
We sincerely thank  Yixi Su,  Werner Schweika, Nikolaos Biniskos, and Markus Grüninger for fruitful discussions on experimental issues resolving the spin splitting and Irene Aguilera for fruitful discussions about the Goldstone theorem in antiferromagnets. This work was supported by several funding sources, including the Deutsche Forschungsgemeinschaft (DFG) through CRC/TRR 227, CRC 1238 (Project C01), TRR 173/3 $-$ 268565370 (Project A11), and TRR 288/2 $-$ 422213477 (Project B06), the European Union (EFRE, Grant No. ZS/2016/06/79307), and the Federal Ministry of Education and Research of Germany (BMBF) within the framework of the Palestinian-German Science Bridge (BMBF Grant No. DBP01436).
\end{acknowledgments}



%

\end{document}